\documentclass[twocolumn,showpacs,preprintnumbers,amsmath,amssymb,aps]{revtex4-1}
\usepackage{graphicx}
\usepackage{epstopdf}           
\usepackage{dcolumn}
\usepackage{bm}
\usepackage[T1]{fontenc} 
\usepackage{lmodern}
\usepackage{braket}
\usepackage{ulem}
\usepackage{color}

\usepackage{hyperref}

\usepackage{braket}

\begin{document}

\title{Magneto-optics of excitons interacting with magnetic ions in CdSe/CdMnS colloidal nanoplatelets}

\author{E.~V.~Shornikova$^1$, D.~R.~Yakovlev$^{1,2}$, D.~O.~Tolmachev$^{1,2}$, V.~Yu.~Ivanov$^3$, I.~V.~Kalitukha$^2$, V.~F.~Sapega$^2$, D.~Kudlacik$^1$,  Yu.~G.~Kusrayev$^2$, A.~A. Golovatenko$^{2}$, S. Shendre$^{4}$, S.~Delikanli$^{4,5}$, H. V.~Demir$^{4,5}$, and M.~Bayer$^{1,2}$}

\affiliation{
$^1$Experimentelle Physik 2, Technische Universit\"at Dortmund,~44227 Dortmund, Germany \\
$^2$Ioffe Institute, Russian Academy of Sciences, 194021 St. Petersburg, Russia \\
$^3$Institute of Physics, Polish Academy of Sciences, PL-02-668 Warsaw, Poland \\
$^4$LUMINOUS! Center of Excellence for Semiconductor Lighting and Displays, School of Electrical and Electronic Engineering, School of Physical and Materials Sciences, Nanyang Technological University, 639798 Singapore\\
$^5$Department of Electrical and Electronics Engineering, Department of Physics, UNAM -- Institute of Materials Science and Nanotechnology, Bilkent University, 06800 Ankara, Turkey
}

\date{\today}

\begin{abstract}
Excitons in diluted magnetic semiconductors represent excellent probes for studying the magnetic properties of these materials. Various magneto-optical effects, which depend sensitively on the exchange interaction of the excitons with the localized spins of the magnetic ions can be used for probing. Here, we study core/shell CdSe/(Cd,Mn)S colloidal nanoplatelets hosting diluted magnetic semiconductor layers. The inclusion of the magnetic Mn$^{2+}$ ions is evidenced by three magneto-optical techniques using high magnetic fields up to 15~T: polarized photoluminescence, optically detected magnetic resonance, and spin-flip Raman scattering. In particular, information on the Mn$^{2+}$ concentration in the CdS shell layers can be obtained from the spin-lattice relaxation dynamics of the Mn$^{2+}$ spin system.
\end{abstract}

\maketitle


\includegraphics{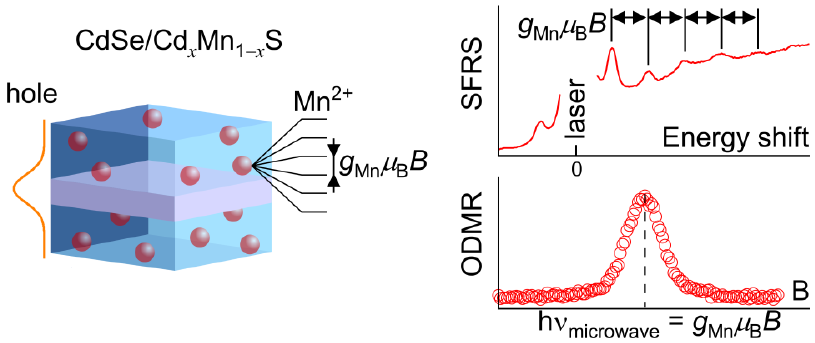}

Incorporation of magnetic ions in colloidal nanocrystals (NCs) opens exciting opportunities for engineering of spintronics devices.~\cite{Efros2001,Beaulac2008afn,Muckel2017,Moro2019} The underlying idea to exploit the strong $sp-d$ exchange interactions between electrons and holes with the localized spins of magnetic ions originates from the physics of diluted magnetic semiconductors (DMSs).~\cite{Furdyna1988} This research direction was established first for bulk DMS materials, and later was successfully extended for epitaxially grown DMS heterostructures, including quantum wells and quantum dots.~\cite{Furdyna1988book,Kossut2010} In colloidal nanostructures it is still at an early stage, while several important results have been already achieved. The giant Zeeman splitting has been demonstrated by measuring the magnetic circular dicroism~\cite{Hoffman2000,Norris2001,Archer2007,Bussian2009}, including the photoinduced magnetism in Ag$^+$-doped CdSe NCs \cite{Pinchetti2018}, and evidenced by polarized photoluminescence in external magnetic fields.~\cite{Beaulac2008nl-1,Long2012,Turyanska2014,Delikanli2015,Murphy2016,Najafi2020} The exchange interaction of excitons with Mn$^{2+}$ ions was proven by optically detected magnetic resonance (ODMR).~\cite{Strassberg2019} Magnetic polaron formation was reported~\cite{Beaulac2009,Rice2017,Muckel2017,Lorenz2020}, and the influence of Mn$^{2+}$ spin fluctuations was considered.~\cite{Rice2016}

Magneto-optical studies of the exciton emission, its giant Zeeman splitting and polarization, are a valuable tool for investigation of DMS nanostructures. There is, however, a limitation for the parameters of DMS NCs to provide efficient exciton photoluminescence. The Mn$^{2+}$ ions have an absorption band at the energy of about 2.1~eV associated with the internal transition $^6A_1\rightarrow^4T_1$; the corresponding $^4T_1\rightarrow^6A_1$ emission is located at about 2.0~eV. This means that in (Cd,Mn)Se NCs the exciton resonance should be considerably detuned from this energy, because the efficient energy transfer to the Mn$^{2+}$ ions would otherwise represent a nonradiative recombination channel for the excitons, quenching their emission. For this reason, (Cd,Mn)Se spherical NCs with large diameters were synthesized in order to keep the exciton emission energy below 2.1~eV.~\cite{Beaulac2008nl-1,Bussian2009,Rice2017}

Nanoplatelets (NPLs) are an emerging class of colloidal nanocrystals, which are atomically flat with several monolayer thickness, resembling free-standing quantum wells.~\cite{Ithurria2008} NPLs with magnetic Mn$^{2+}$ ions were synthesized in 2015~\cite{Delikanli2015}, providing remarkable opportunities for wave-function engineering.~\cite{Bussian2009,Furdyna2010ch4,Muckel2018,Zhang2019} The Mn$^{2+}$ ions were incorporated in the NPL cores~\cite{Davis2019} or shells.~\cite{Delikanli2015,Muckel2018}

In this paper, we study the magneto-optical properties of core/shell CdSe/Cd$_{1-x}$Mn$_x$S NPLs, which arise from excitons interacting with the magnetic ions. Three experimental approaches are used for that: (i) polarized photoluminescence in external magnetic fields, (ii) optically detected magnetic resonance of the Mn$^{2+}$ ions, and (iii) spin-flip Raman scattering.  We measure the spin-lattice relaxation dynamics of the Mn$^{2+}$ spin system and suggest an approach for evaluation of the Mn$^{2+}$ concentration. 

\section{Samples}
Four NPL samples were grown for this study, see Refs. \onlinecite{Delikanli2015,Delikanli2019,Shendre2019} and Supplementary Section~S4 for details. All of them have a 2-monolayer thick CdSe core and 4-monolayer thick shells cladding the core. The reference sample \#0 has nonmagnetic CdS shells and the other three DMS NPLs have Cd$_{1-x}$Mn$_x$S shells with Mn$^{2+}$ concentrations $x$ ranging from 0.009 to 0.029. The sample parameters are given in Table~\ref{tab1}. Note that the nominal Mn$^{2+}$ concentrations obtained by inductively coupled plasma mass spectrometry (ICP-MS) measurements differ from the values that we evaluate from the spin-lattice relaxation dynamics. We are convinced that the latter values are more reliable so that we use them in the paper.   

\begin{table}
	\scriptsize
	\caption{Parameters of the studied CdSe/CdS and CdSe/Cd$_{1-x}$Mn$_x$S NPLs. Mn$^{2+}$ concentrations, nominal (measured by ICP-MS) and evaluated from the spin-lattice-relaxation dynamics in ODMR experiments.}
	\begin{tabular}{ |c|c|c|c|c| } 
		\hline
		Sample & \parbox[c]{2cm}{Nominal Mn$^{2+}$ content from ICP-MS} & \parbox[c]{0.8cm}{$\tau_{\rm SLR}$, $\mu$s}   & \parbox[c]{2cm}{Evaluated Mn$^{2+}$ content from ODMR}&
		\parbox[c]{0.9cm}{$\Delta E_{\rm AF}$, meV}  \\ 
		\hline
		\#0 & 0 & -- & --&1.7\\
		\#1 & 0.012 & 405 & 0.009&1.6\\ 
		\#2 & 0.019 & 350 & 0.010&1.8\\ 
		\#3 & 0.050 & 20 & 0.029&1.9\\
		\hline
	\end{tabular}
	\label{tab1}
\end{table}

\section{Specifics of DMS heterostructures} 
The band structure of the CdSe/Cd$_{1-x}$Mn$_x$S NPLs is shown schematically in Figure~\ref{fig2}a. The CdSe core with cubic lattice has the bandgap energy $E_{\rm g}^{\rm CdSe}=1.75$~eV~\cite{Adachi2004}, and is sandwiched between shells with $E_{\rm g}^{\rm CdS}=2.50$~eV. Note that the $E_{\rm g}^{\rm CdSe}=1.84$ eV used in Refs.~\onlinecite{Strassberg2019,Muckel2018} corresponds to the wurtzite lattice. The conduction and valence band offsets between CdSe and CdS are not precisely known. However, the valence band offsets reported in literature are large, so that the hole is believed to be confined in the CdSe core. The reported conduction band offsets range from $300$~meV to 0~meV, and this value depends on the crystal structure, NC size, lattice strain and temperature. Due to the quite weak, if present at all, confinement, the electron wave function leaks into the CdS shell (for references, see~\cite{Javaux2013}). The electron and hole wave functions are centered in the nonmagnetic CdSe core and only partially penetrate into the DMS shell. For this reason, all exchange effects in the studied DMS NPLs are expected to be reduced compared to bulk DMSs with the same Mn$^{2+}$ concentration. Quantum mechanical calculations give an estimate of the electron wave function fraction in the shell of 60\% and the hole fraction of 30\%, see Supplementary Section~S2.

There are several factors that need to be taken into account for evaluation of the strength of the exciton and carrier exchange interactions with the Mn$^{2+}$ spins in CdSe/Cd$_{1-x}$Mn$_x$S NPLs:  
\begin{itemize}
	\item [(i)] Penetration of the electron and hole wave functions into the DMS shells. Note that in bulk II-VI DMSs the exchange interaction of holes is $4-5$ times stronger than that of the electrons. Correspondingly, the holes provide the dominating contributions to the magneto-optical effects, like the giant Zeeman splitting of exciton states, Faraday rotation, formation of exciton magnetic polarons, etc. 
	\item [(ii)] Modification of the electron exchange constant $\alpha$ by strong quantization. $\alpha$ is reduced with increasing confinement and can even change its sign.~\cite{Merkulov1999,Merkulov2010ch3}  
	\item [(iii)] Variations of the magnetic properties of the Mn$^{2+}$ spin system, which are controlled by the Mn$^{2+}$--Mn$^{2+}$ interactions and are different in bulk DMSs and in thin DMS layers or at the interfaces between DMS and nonmagnetic layers, because of different statistics of neighboring Mn$^{2+}$ spins.~\cite{Yakovlev2010ch8}
\end{itemize}

Therefore, it is a challenging task to account properly for the contributions of these factors to the magneto-optical properties of the studied NPLs. As a result, one can not use most of the established approaches in DMS physics for evaluation of the Mn$^{2+}$ content by means of magneto-optical techniques.

\section{Time-integrated and time-resolved photoluminescence} 
\begin{figure*}[t!]
	\begin{center}
		\includegraphics{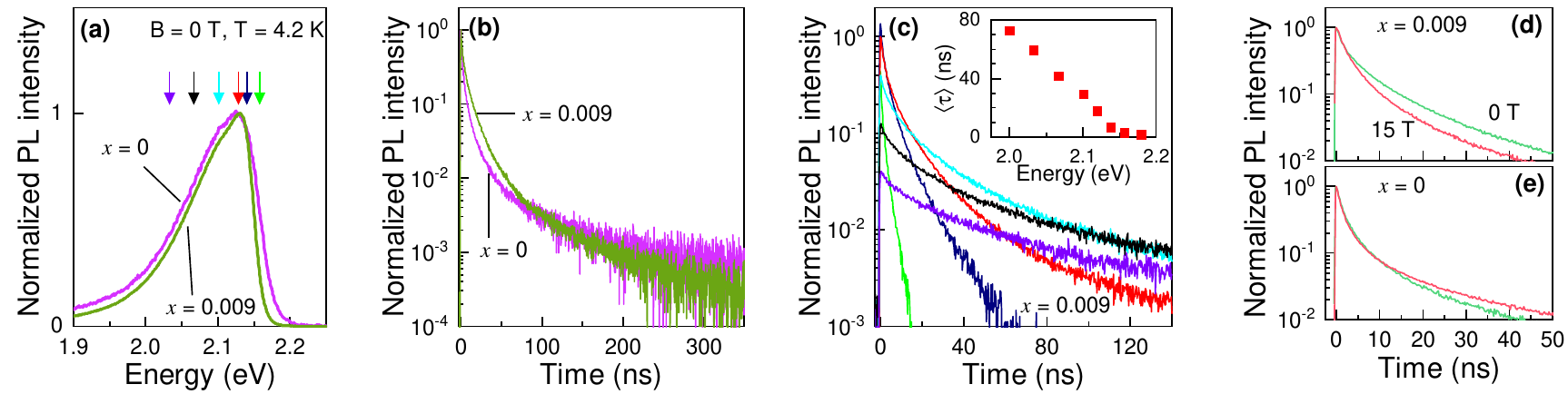}
		\caption{\label{fig1} (a) Photoluminescence spectra of CdSe/CdS (Sample \#0, pink) and CdSe/Cd$_{0.991}$Mn$_{0.009}$S (Sample \#1, green) NPLs at $T=4.2$~K, $B=0$~T. (b) Photoluminescence decay traces of CdSe/CdS (Sample \#0, pink) and CdSe/Cd$_{0.991}$Mn$_{0.009}$S (Sample \#1, green) NPLs, measured at their PL maxima of 2.127~eV. (c) Photoluminescence decay curves of CdSe/Cd$_{0.991}$Mn$_{0.009}$S  NPLs (Sample \#1) at various detection energies shown by the arrows in panel (a) using the same color code. Inset: Spectral dependence of the average  decay time $\langle\tau\rangle$ in Sample \#1.    (d,e) PL decay at 2.127~eV measured in magnetic fields $B=0$~T (green) and  15~T (red) in Samples \#1 and \#0.}   
	\end{center}
\end{figure*}

\begin{figure*}[t!]
	\begin{center}
		\includegraphics{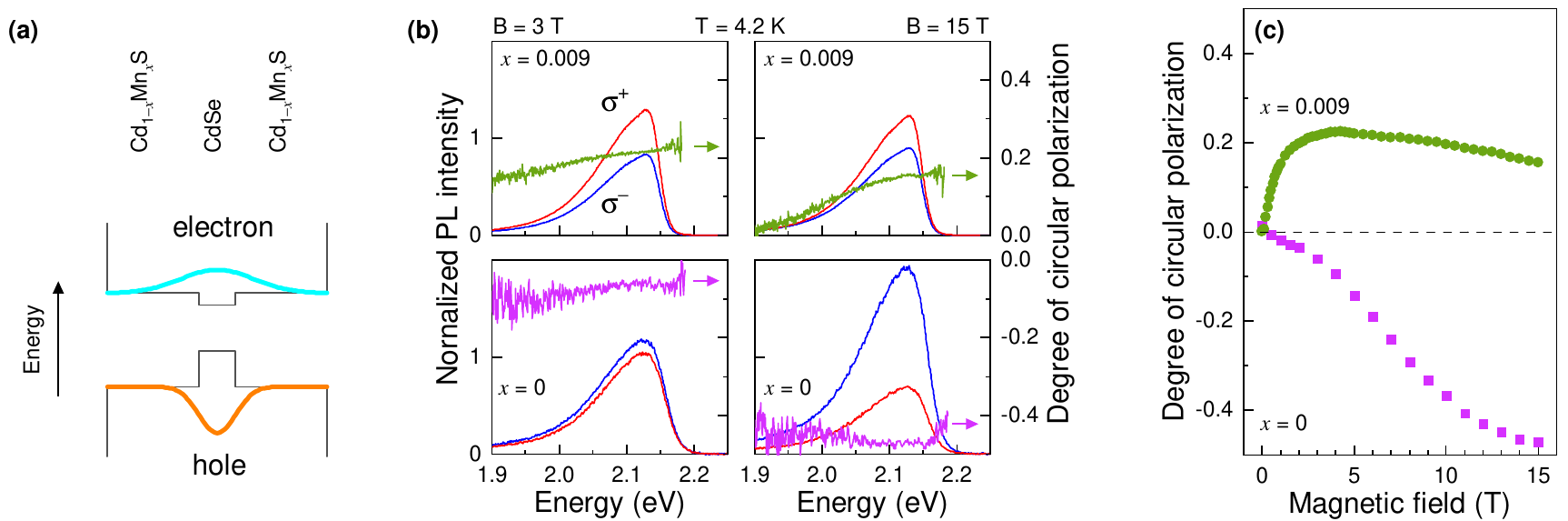}
		\caption{\label{fig2} (a) Schematic diagram of the electron and hole wave functions (cyan and orange lines) in CdSe/Cd$_{1-x}$Mn$_x$S NPLs. (b) Photoluminescence spectra of the $\sigma^+$ (red) and $\sigma^-$ (blue) polarized components in CdSe/CdS NPLs (Sample \#0, bottom) and CdSe/Cd$_{0.991}$Mn$_{0.009}$S NPLs (Sample \#1, top) at $T=4.2$~K, $B=3$~T (left) and 15~T (right). (c) Magnetic field dependence of the DCP in Sample \#0 (pink) and Sample \#1 (green). } 
	\end{center}
\end{figure*}

Figure~\ref{fig1}a shows photoluminescence (PL) spectra of CdSe/CdS (Sample \#0) and  CdSe/Cd$_{0.991}$Mn$_{0.009}$S (Sample \#1) NPLs. Both spectra are very similar to each other, so that implementation of a small Mn$^{2+}$ concentration does not change the PL. The emission lines of both samples are centered at $2.127$~eV (red arrow) and have full widths at half maximum of about 100~meV, which is typical for core/shell NPLs.~\cite{Shornikova2018nl} As the PL is close to the $^4T_1\rightarrow^6A_1$ transition of the Mn$^{2+}$ ions at 2.1~eV, our first task is to identify the origin of the emission from CdSe/Cd$_{0.991}$Mn$_{0.009}$S NPLs and to prove that it is dominated by exciton recombination. The similarity of the PL spectra of nonmagnetic and DMS NPLs gives a first hint for that. 

The recombination dynamics can be also used for the identification of the origin of the emission. It is known that at liquid helium temperatures  the decay of the Mn$^{2+}$ emission via the $^4T_1\rightarrow^6A_1$ transition is very slow, occurring on time scales in the $10-500$~$\mu$s range in bulk DMSs,~\cite{Mueller1986,Schenk1996} like (Cd,Mn)Te, (Zn,Mn)Te and (Zn,Mn)S, and is 270~$\mu$s in (Cd,Mn)Se colloidal quantum dots.~\cite{Beaulac2008nl-2} The exciton recombination dynamics is by a few orders of magnitude faster, happening for neutral excitons in the range of 1~ns to 1~$~\mu$s, depending on relative involvement of bright and dark exciton states, or of a few nanoseconds for charged excitons (trions).~\cite{Shornikova2018nl,Shornikova2018ns,Shornikova2020nl}

The time-resolved recombination dynamics measured at the PL maxima in Samples \#0 and \#1 are shown in Figure~\ref{fig1}b. In both cases, the decay of the PL intensity is faster
more than three orders of magnitude compared to that of Mn$^{2+}$ emission and takes place within 300 ns. This allows us to conclude that the dominating part of the emission in CdSe/Cd$_{0.991}$Mn$_{0.009}$S NPLs is provided by exciton recombination and the Mn$^{2+}$ emission is very weak, if present at all. The two other DMS samples have similar properties.    

As it is common for the colloidal nanocrystals, the recombination dynamics in the studied NPLs do not show a monoexponential decay. For example, the decays at the PL maxima, shown in Figure~\ref{fig1}b, require a fit with a three-term exponential function for a good description: $I(t) = A_1\exp(-t/\tau_1)+A_2\exp(-t/\tau_2)+A_3\exp(-t/\tau_3)$. The three decay times are  $\tau_1=3$~ns,  $\tau_2=11$~ns and $\tau_3=56$~ns for Sample \#0, and $\tau_1=3$~ns, $\tau_2=12$~ns and $\tau_3=45$~ns for Sample \#1. Note that they are close to each other in these nonmagnetic and DMS NPLs. 

The spectral dependence of the PL dynamics in  CdSe/Cd$_{0.991}$Mn$_{0.009}$S NPLs is given in Figure~\ref{fig1}c. The general trend is that the decay times increase with decreasing emission energy. This is clearly seen in the inset of Figure~\ref{fig1}c, where the spectral dependence of the average decay time $\langle\tau\rangle$ is given. $\langle\tau\rangle$ is calculated as  $\langle\tau\rangle=\tau_1\nu_1+\tau_2\nu_2+\tau_3\nu_3$, where $\nu_i=A_i\tau_i/(A_1\tau_1+A_2\tau_2+A_3\tau_3)$. The average decay time increases from 2 up to 70 ns from the high- to the low-energy tail. Such behavior is typical for ensembles of colloidal nanocrystals with an efficient F\"orster energy transfer.~\cite{Furis2005,Liu2015}

Further more, the recombination dynamics are weakly affected by external magnetic fields. This is shown in Figures~\ref{fig1}d,e, where the PL decays at the emission maximum are compared for $B=0$ and 15~T.

Note that the character of the recombination dynamics at low temperatures in colloidal NPLs and its dependence on magnetic field allows one to identify whether the emission is contributed by neutral or by charged excitons.~\cite{Shornikova2018nl,Shornikova2018ns,Shornikova2020nl} For example, at $T=4.2$~K the trion emission in CdSe/CdS NPLs with thick shells is monoexponential with a decay time of 3~ns and is independent of magnetic field. Contrary to that, the decay of neutral excitons in CdSe NPLs has a bi-exponential decay with a very fast initial component of 20~ps and a long component of 80~ns, which shortens with increasing magnetic field. 

The recombination dynamics in the NPLs studied in this paper do not clearly correspond to either neutral or charged exciton behavior, but are rather superpositions of both. Additionally, for resonant excitation, we clearly observe emission from dark excitons (Figure~\ref{fig_SFRS_1}a). The bright-dark exciton energy splitting, $\Delta E_{\rm AF}$, ranges between 1.6 and 1.9~meV (see below). Therefore, at $T=4.2$~K the bright state has about 1\% population in thermal equilibrium, and should contribute to the emission. We also detect electron spin flips, which means that some of the NPLs are negatively charged, i.e. they may contain negatively charged excitons (Figure~\ref{fig_SFRS_1}a). From all these findings, we conclude that the PL is contributed by recombination of neutral (bright and dark) and charged excitons. More details are given in the Supplementary Section~S1.

\section{Polarized photoluminescence in magnetic field} 
This technique, which exploits the exciton (trion) spin polarization on their Zeeman split sublevels, is a sensitive tool to measure small spin splittings comparable with the thermal energy $k_{\rm B}T$, where $k_{\rm B}$ is Boltzmann constant.~\cite{Liu2013,Shornikova2018nl,Shornikova2020nn}
The experiment is relatively easy in realization, but requires liquid helium temperatures and strong magnetic fields of $10-15$~T or even up to $30-65$~T to gather sufficient information on the linear dependence of the circular polarization degree of PL, $P_{\rm c}(B)$, on magnetic field for rather weak fields, until it reaches saturation in high magnetic fields, $P^{sat}_{\rm c}$.    
The degree of circular polarization (DCP) is defined as $P_{\rm c}=(I^+-I^-)/(I^++I^-)$, where $I^+$ and $I^-$ are the PL intensities of the $\sigma^+$ and $\sigma^-$ circularly polarized components, respectively. The magnetic field is applied in the Faraday geometry, i.e., parallel to the emission wave vector direction.

Figure~\ref{fig2}b shows polarized PL spectra of the nonmagnetic Sample \#0 (bottom) and Mn-doped Sample \#1 (top), measured in  magnetic fields $B=3$ and 15~T. One can clearly see the difference between DCP in nonmagnetic and DMS NPLs. First, they have opposite signs. In CdSe/CdS NPLs $P_{\rm c}<0$, its absolute value increases with growing magnetic field about monotonically and saturates above 12~T (Figure~\ref{fig2}c). At $B=15$~T it reaches $P_{\rm c}=-0.47$. This behavior is similar to what was reported for thick-shell CdSe/CdS NPLs (see Figure~3c in Ref.~\onlinecite{Shornikova2018nl}), where the emission was provided by negatively charged excitons.   

In the DMS sample $P_{\rm c}>0$, it increases fast, reaching the plateau value of $+0.22$ at $B=4$~T, and then slowly decreases in higher magnetic fields. The $P_{\rm c}$ sign reversal in II--VI DMS materials, compared to the nonmagnetic reference, is a clear evidence of the exchange interaction of charge carriers with the Mn$^{2+}$ ions. It is provided by the signs of the exchange constants in the conduction ($\alpha>0$) and valence bands ($\beta<0$).~\cite{Kossut2010} More details are given in the Supplementary Section~S3. 

\section{Analysis of polarized photoluminescence} The exciton and trion DCP can be written as 
\begin{equation} \label{Pc1}
	P_{\rm c}(B)= -P^{sat}_{\rm c} \frac{\tau}{\tau+\tau_{s}} \tanh  \frac{\Delta E_{\rm Z}(B)}{2k_{\rm B}T} \, .
\end{equation} 
Here $\Delta E_{\rm Z}(B)$ is the Zeeman splitting, $\tau$ is the lifetime and $\tau_{s}$ is the spin relaxation time, and $P^{sat}_{\rm c}$ is the saturation degree of polarization, which depends on the specifics of the spin level structure and NPL orientation in the ensemble. 

In nonmagnetic samples the intrinsic exciton Zeeman splitting is
\begin{equation} \label{Pc2}
	\Delta E_{\rm Z,X}(B)= g_{\rm X}\mu_{\rm B}B\, ,
\end{equation} 
where $g_{\rm X}$ is the exciton $g$-factor and $\mu_{\rm B}$ is the Bohr magneton. Accounting for the specifics of the bright and dark excitons is considered in the Supplementary Section~S3.

In DMS samples an additional term, $E_{\rm exch,X}(B)$, describing the exciton exchange interaction with the Mn$^{2+}$ spins has to be added 
\begin{equation} \label{Pc3}
	\Delta E_{\rm Z,X}(B)= g_{\rm X}\mu_{\rm B}B+E_{\rm exch,X}(B)\, .
\end{equation}
Note that $E_{\rm exch,X}(B)$ is controlled by the exchange interaction of both electron and hole composing the exciton with the Mn$^{2+}$ ions, and therefore depends on the overlap of the electron and hole wave functions with the (Cd,Mn)S shells. 

For the negative trion, being composed of two electrons and one hole, the Zeeman splitting is determined by the hole splitting:
\begin{equation} \label{Pc2T}
	\Delta E_{\rm Z,h}(B)= -3g_{\rm h}\mu_{\rm B}B\, ,
\end{equation} 
where $g_{\rm h}$ is the hole $g$-factor. In DMS samples 
\begin{equation} \label{Pc3T}
	\Delta E_{\rm Z,h}(B)= -3g_{\rm h}\mu_{\rm B}B+E_{\rm exch,h}(B)\, .
\end{equation}
$E_{\rm exch,h}(B)$ is determined by the exchange interaction of the hole with the Mn$^{2+}$ ions. 
Here we use the definition of the hole $g$-factor sign that is commonly used for colloidal nanocrystals.~\cite{Efros2003,Shornikova2018nl}  In the frame of this convention, the intrinsic hole Zeeman splitting provided by the negative hole $g$-factor ($g_{\rm h}<0$) is in competition with the hole exchange splitting determined by $\beta<0$. On the other hand, for the conduction band electron, both the intrinsic Zeeman splitting ($g_{\rm e}>0$) and the exchange one with $\alpha>0$ add to each other.  

The electron $g$-factor is $g_{\rm e}=+1.70$ in CdSe/CdS NPLs (see below). The hole $g$-factor $g_{\rm h}=-0.7$ was measured in high magnetic fields.~\cite{Shornikova2018nl} For small $\Delta E_{\rm AF}$, as in our case, the $g$-factor of the bright exciton is $g_{\rm XA}=-g_{\rm e}-3g_{\rm h}=+0.4$. This value matches well with $g_{\rm XA}=+0.32$ measured in 4 monolayer thick bare core NPLs by absorption spectroscopy in high magnetic fields \cite{Brumberg2019}. For  the dark exciton $g_{\rm XF}=g_{\rm e}-3g_{\rm h}=+3.8$. One can see from Eq.~\eqref{Pc1} that the negative DCP found in experiment requires $g_{\rm X}>0$, i.e. can be achieved for the bright and the dark excitons. In case of the negative trion, $P_{\rm c}<0$ requires $g_{\rm h}<0$, see Eqs.~\eqref{Pc1} and \eqref{Pc2T}, which is indeed the case for CdSe/CdS NPLs. To summarize, the negative DCP observed in nonmagnetic CdSe/CdS NPLs can be provided by the dark and bright excitons and the negative trions.

In DMS NPLs the polarization is positive, which requires a negative sign of the Zeeman splitting $\Delta E_{\rm Z}$. One can see from Eqs.~\eqref{Pc3} and \eqref{Pc3T} that this can be the case when the intrinsic and exchanges terms have different signs as well as when for the exciton case $|E_{\rm exch,X}(B)| > |g_{\rm X}\mu_{\rm B}B|$ and for the trion case $|E_{\rm exch,h}(B)| > |3g_{\rm h}\mu_{\rm B}B|$. In the trion case, the fulfillment of this condition is solely provided by the hole exchange interaction with the Mn$^{2+}$ ions, i.e. requires a sufficiently large penetration of the hole wave function into the DMS shells. In CdSe, $g_{\rm e}$ and $\alpha$ are both positive and, therefore, the Zeeman splitting for the conduction band electron can not be inverted. As a result, for the exciton case the inversion of the DCP sign can also be provided only by the hole exchange with the Mn$^{2+}$ spins. To support this conclusion, we provide in the Supplementary Section~S3 results of model calculations for the bright and dark excitons and for the negative trions for various penetrations of the hole wave functions into the DMS shells.

\section{Optically detected magnetic resonance (ODMR)} The ODMR technique combines the advantage of resonant excitation of spin states by microwave radiation with the high sensitivity of optical detection of the induced changes. It is especially useful for the investigation of semiconductor nanostructures, whose small volume is not sufficient to provide sufficiently strong signals for the electron paramagnetic resonance technique. Additionally, the possibility of spectrally selective detection on specific optical resonances, e.g., the exciton or impurity related emission, allows one to obtain a clear identification of the addressed electronic transitions.

\begin{figure*}[t!]
	\includegraphics* {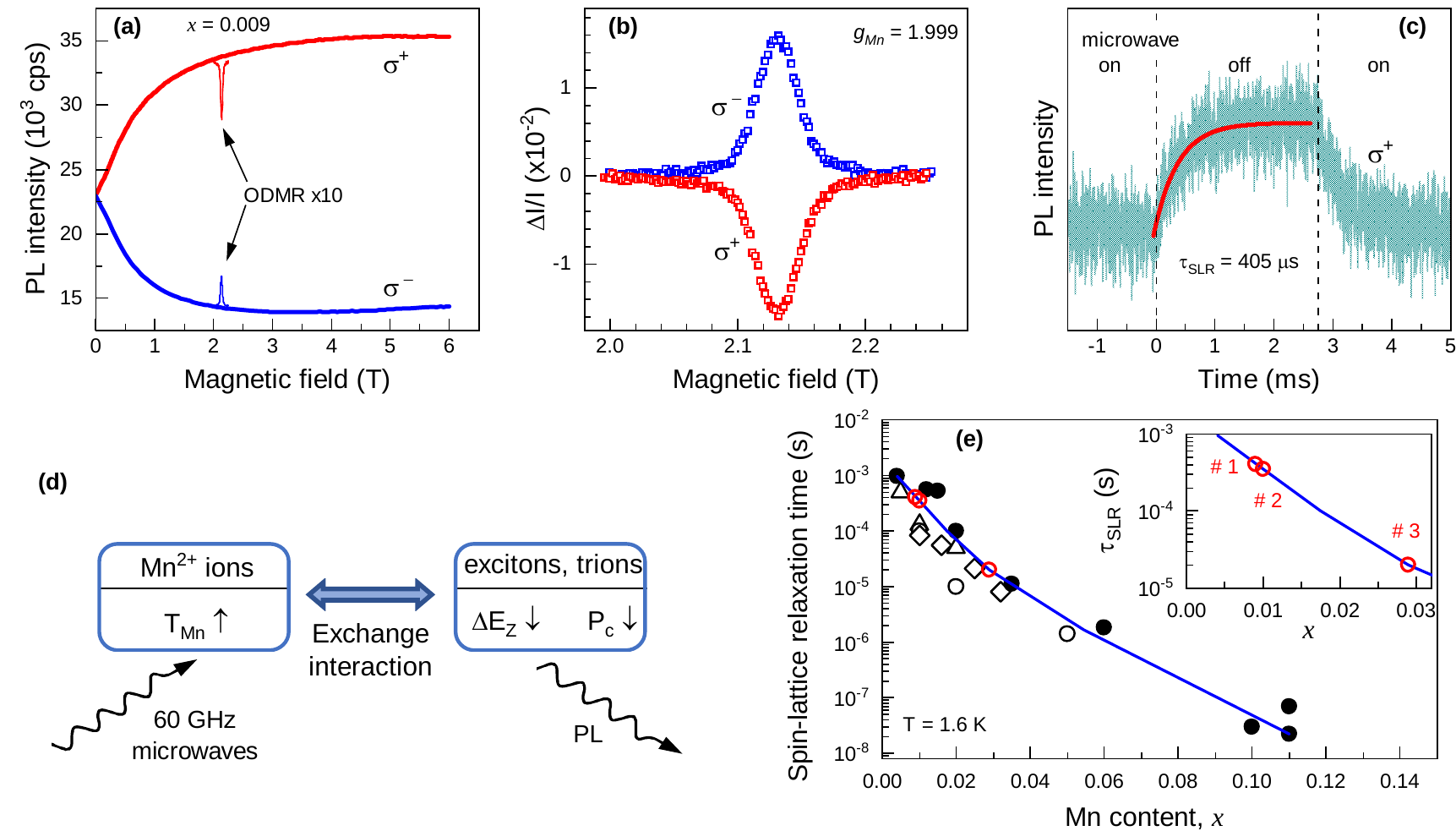}
	\caption{(a)  Dependence of the $\sigma^+$ (red) and $\sigma^-$ (blue) circularly polarized components of photoluminescence on magnetic field in Faraday geometry for CdSe/Cd$_{0.991}$Mn$_{0.009}$S NPLs (Sample \#1). $P_{exc}=4$~W/cm$^2$ and $T=1.8$~K. The peaks show the PL intensities for application of microwaves (ODMR signals). Their intensities are multiplied by a factor of 10. (b) Normalized ODMR signals for the two circularly polarized components of the PL. The ODMR resonance has a linewidth of $\Delta B \approx 40$~mT and is located at $B = 2.130$~T, which corresponds to a $g$-factor of 1.999. The $\sigma^+$ component decreases at resonance conditions (red), while the $\sigma^-$ component increases (blue). (c) Temporal evolution of the $\sigma^+$ PL component after switching on/off the microwaves for resonance excitation at $B = 2.130$~T. The red line is an exponential fit with a time constant $\tau_{\rm SLR}=405$~$\mu$s. (d) Schematic diagram of the interactions between the Mn$^{2+}$ and exciton (trion) spin systems. (e) Spin-lattice relaxation time as a function of the Mn$^{2+}$ content $x$; adapted from Ref.~\onlinecite{Yakovlev2010ch8}. Blue line is a guide for the eye, red circles are experimental data measured in the present work (see Table~\ref{tab1}).}
	\label{Fig3}
\end{figure*}

In the case of diluted magnetic semiconductors, the resonant microwave heating of the Mn$^{2+}$ ions increases the Mn$^{2+}$ spin temperature $T_{{\rm Mn}}$ and, consequently, reduces the Mn$^{2+}$ spin polarization $\langle S_{\rm Mn} \rangle$. These changes can be detected optically via the excitons or trions interacting with the Mn$^{2+}$ spins, see Figure~\ref{Fig3}d. The application of the ODMR technique to  quantum well structures based on (Zn,Mn)Se DMS was discussed in Refs.~\onlinecite{Ivanov2007,Ivanov2008,Yakovlev2010ch8}. There it was shown that the resonant heating of the Mn$^{2+}$ spin system can be detected by several effects: (i) the decrease of the exciton giant Zeeman splitting, resulting in a spectral shift of the exciton emission line, (ii) the decrease of the circular polarization degree induced by the magnetic field, and (iii) the redistribution of the emission intensity between the exciton line and the Mn$^{2+}$ emission band. Recently, ODMR measured at 10~GHz microwave radiation via polarized photoluminescence was reported for CdSe/Cd$_{1-x}$Mn$_x$S NPLs.~\cite{Strassberg2019} 

Figure~\ref{Fig3}a shows the intensities of the $\sigma^+$ and $\sigma^-$ PL components of the CdSe/Cd$_{0.991}$Mn$_{0.009}$S NPLs (Sample \#1), measured versus magnetic field without and with microwaves. As discussed above, the $\sigma^+$ component has a higher intensity due to the stronger thermal population of the excitons (trions) on the associated Zeeman sublevels split in magnetic field. Without microwaves, the intensities of these components change smoothly with magnetic field, following the DCP trend shown in Figure~\ref{fig2}c. In the presence of 59.6~GHz microwave radiation, two sharp resonances are observed at $B = 2.130$~T. The resonant decrease of the $\sigma^+$ intensity and the correlated increase of the $\sigma^-$ intensity evidence heating of the Mn$^{2+}$ spin system, which accordingly decreases the exciton (trion) giant Zeeman splitting and the exciton (trion) DCP.~\cite{Ivanov2008,Yakovlev2010ch8} The PL intensity variations normalized to the relative PL intensities without microwaves ($I$), are shown in more detail in Figure~\ref{Fig3}b. They represent broad peaks with a width of $\Delta B=40$~mT and are centered at $B = 2.130$~T corresponding to a $g$-factor of $1.999\pm0.005$. This $g$-factor matches with the Mn$^{2+}$ value of $g_{\rm Mn}=2.01$, reported for ZnS:Mn$^{2+}$ and CdTe:Mn$^{2+}$ in electron spin resonance measurements.~\cite{Matarrese1956, Lambe1960}

The spin-lattice relaxation (SLR) dynamics of the Mn$^{2+}$ spin system can be measured by modulating the microwave radiation between on and off and time-resolved detection of the changes induced thereby, reflecting cooling or heating of the Mn$^{2+}$ spins. An example of such a measurement for CdSe/Cd$_{0.991}$Mn$_{0.009}$S NPLs is shown in Figure~\ref{Fig3}c. Here, the red line is an exponential fit with the characteristic spin-lattice relaxation time $\tau_{\rm SLR}=405$~$\mu$s. Similar measurements performed for Samples \#2 and \#3 give 350~$\mu$s and 20~$\mu$s, respectively (Table~\ref{tab1}).

As we discussed above, most of the magneto-optical approaches that are commonly used for evaluation of the Mn$^{2+}$ concentration in bulk DMS, cannot be directly applied to CdSe/Cd$_{1-x}$Mn$_x$S NPLs. We suggest that a quite accurate evaluation can be achieved from the spin-lattice relaxation time $\tau_{\rm SLR}$. It is known that the $\tau_{\rm SLR}$ of the Mn$^{2+}$ ions in II-VI semiconductors has a very strong dependence on the Mn$^{2+}$ concentration, which covers about five orders of magnitude from 1~ms down to 10~ns with increasing $x$ from 0.004 up to 0.11, see Figure~\ref{Fig3}e, where the data from Figure~8.10 in Ref.~\onlinecite{Yakovlev2010ch8} are reproduced.  This strong dependence arises from the quenching of the orbital momentum of the $d$-electrons in the Mn$^{2+}$ ion, i.e., it has zero orbital momentum ($L=0$) and does not interact with the phonon system. The only mechanisms that provide spin-lattice relaxation for the Mn$^{2+}$ ions are given by the Mn$^{2+}$--Mn$^{2+}$ interactions, which obviously are strongly dependent on the number of neighboring Mn$^{2+}$ ions and the distances between them, which in turn strongly change with increasing Mn$^{2+}$ concentration. The red lines in Figure~\ref{Fig3}e mark the times that we measured for the CdSe/Cd$_{1-x}$Mn$_x$S NPLs. Their comparison with the literature data shown by the symbols allows us to evaluate the Mn$^{2+}$ concentration for the studied samples. As one can see from Table~\ref{tab1}, the nominal Mn$^{2+}$ concentration measured by the ICP-MS method is in good agreement with our data for Samples \#1 and \#2, but differs for Sample \#3. We emphasize that the suggested approach for evaluation of the Mn$^{2+}$ concentration is very reliable and could be widely used for nanostructures. 

\begin{figure*}[t]
	\begin{center}
		\includegraphics{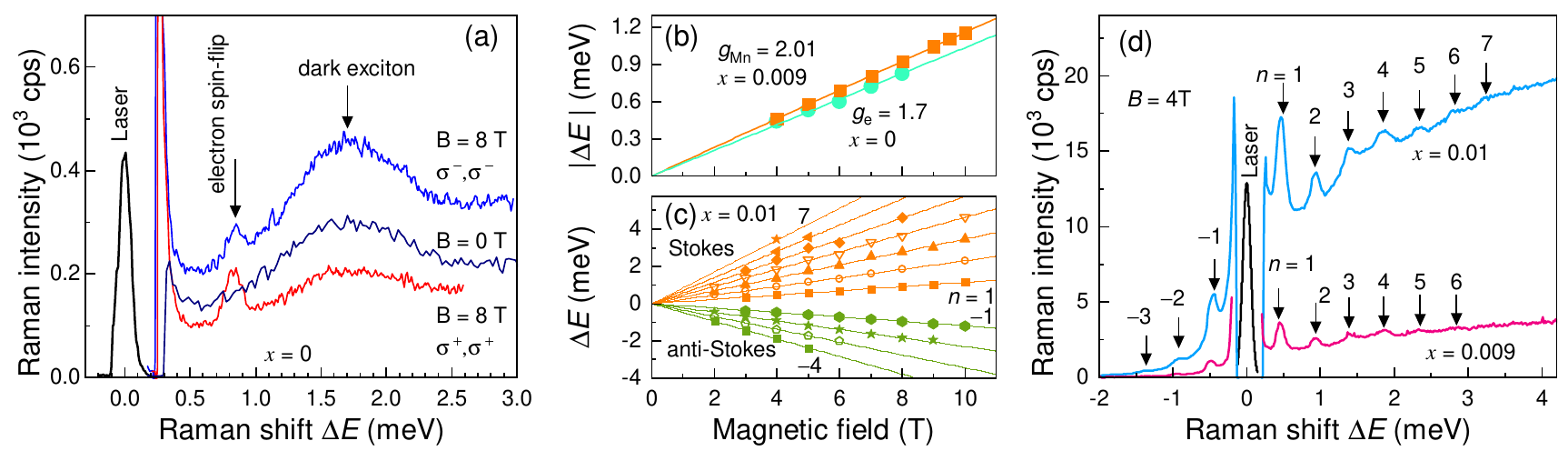}
		\caption{\label{fig_SFRS_1}  (a) Raman spectrum for the circularly co-polarized configuration in Faraday geometry showing the lines of the electron spin-flip and dark exciton for the  nonmagnetic CdSe/CdS NPLs (Sample \#0); $E_{\rm exc} = 2.165$~eV, $P_{\rm exc} = 0.2$~W/cm$^2$, $T=1.6$~K, $B=0$ and 8~T. (b) Magnetic field dependence of the absolute Stokes energy shifts of the $n=1$ Mn$^{2+}$ resonance in CdSe/Cd$_{0.991}$Mn$_{0.009}$S (Sample \#1, orange squares) and of the electron spin flip in CdSe/CdS NPLs (Sample \#0, blue circles) in Faraday geometry. Solid lines are linear fits to the data. (c) Magnetic field dependence of the relative Raman shifts of all detectable Mn$^{2+}$ resonances in CdSe/Cd$_{0.99}$Mn$_{0.01}$S NPLs in Voigt geometry. (d) Mn$^{2+}$ spin scattering spectra for the co-polarized configuration in Faraday geometry for the CdSe/Cd$_{0.991}$Mn$_{0.009}$S  (Sample \#1, pink) CdSe/Cd$_{0.99}$Mn$_{0.01}$S (Sample \#2, blue) NPLs. $E_{\rm exc} = 2.151$~eV, $P_{\rm exc} = 1.6~$W/cm$^2$, $B=4$~T, $T=1.6$~K. In Faraday geometry up to 7 (3) Mn$^{2+}$ resonances are observed in the Stokes (anti-Stokes) spectral ranges.} 
	\end{center}
\end{figure*} 

\section{Spin-flip Raman scattering (SFRS)} The 
SFRS spectroscopy is a sophisticated tool to investigate the Zeeman splittings of carriers (electrons or holes), excitons or magnetic ions. It provides detailed information about their spin structure and spin interactions. The polarization properties of the SFRS lines deliver information on the symmetries of the involved states and allow one to identify responsible flip mechanisms. In SFRS, the Zeeman splitting is obtained from the Raman shift, which is equal to the energy shift between the laser photon energy and the energy of the scattered light. The technique was successfully used for investigation of the exchange interactions of carriers and excitons with the magnetic ions in DMS bulk samples~\cite{Heiman1983,Furdyna1988book} and quantum well structures.~\cite{Stuhler1995,Bao2005,Smith2008} We showed recently that nonmagnetic CdSe and CdSe/CdS NPLs can be studied by SFRS,~\cite{Shornikova2018nl,Kudlacik2020nl} but this technique had not been used so far for DMS colloidal nanocrystals.

Figure~\ref{fig_SFRS_1}a shows Raman spectra of the CdSe/CdS NPLs (Sample \#0) for resonant excitation of the exciton state at $E_{\rm exc} = 2.165$~eV. Here positive values of the Raman shift $\Delta E$ correspond to a Stokes shift of the scattered photons to lower energies. At zero magnetic field there is a relatively broad line with a full width at half maximum (FWHM) of 0.5~meV, whose maximum is shifted by 1.7~meV. This shift does not change in the applied magnetic field of $B=8$~T. We assign it to the energy splitting between the bright and dark exciton states $\Delta E_{\rm AF}$.~\cite{Shornikova2018ns} The dark exciton line was observed in all studied samples with $\Delta E_{\rm AF}$ ranging between 1.6 and 1.9~meV (Table~\ref{tab1}). This supports our assumption that the dark excitons also contribute to the emission from the CdSe/Cd$_{1-x}$Mn$_x$S NPLs.

At $B=8$~T a narrow line associated with the electron spin-flip and shifted by 0.82~meV from the laser is seen in the Raman spectrum (Figure~\ref{fig_SFRS_1}a). Its shift depends linearly on magnetic field (Figure~\ref{fig_SFRS_1}b, blue circles). A fit with $|\Delta E|=|g_{\rm e}|\mu_{\rm B}B$ gives the $g$-factor value $|g_{\rm e}|=1.70\pm0.02$, which is close to the electron $g$-factor measured in CdSe and CdSe/CdS NPLs~\cite{Shornikova2018nl,Kudlacik2020nl}. Note that $g_{\rm e}>0$ in bulk CdSe and in these structures. 

Figure~\ref{fig_SFRS_1}d shows Raman spectra of CdSe/Cd$_{0.991}$Mn$_{0.009}$S and CdSe/Cd$_{0.99}$Mn$_{0.01}$S NPLs measured at $B=4$~T with resonant excitation of the exciton at $E_{\rm exc}=2.151$~eV, $T=1.6$~K. They are obviously very different compared to the spectra in Figure~\ref{fig_SFRS_1}a. No electron spin-flip is detected; instead, a set of equidistant lines is observed. Up to seven lines in the Stokes and up to three lines in the anti-Stokes energy range can be resolved. All these lines shift linearly with magnetic field (see Figure~\ref{fig_SFRS_1}c), following the equation $\Delta E = n g_{\rm Mn}\mu_{\rm B} B$, where $n$ is an integer number. An accurate evaluation of $g_{\rm Mn}=2.01\pm0.03$ is obtained from the fit of the line with $n=1$, shown in Figure~\ref{fig_SFRS_1}b by orange squares. This $g$-factor matches well with the Mn$^{2+}$ $g$-factor of $2.01$~\cite{Matarrese1956, Lambe1960}. Hence, we conclude that the measured Raman signals have to be attributed to spin flips of the Mn$^{2+}$ ions interacting with the photogenerated exciton. We measured the spectral dependence of the Raman signal intensities. The maximal signal is reached when the laser is in resonance with the exciton. This shows that the exciton serves as an intermediate scattering state, which resonantly enhances the Raman cross-section.

The Raman signal is detected in all four combinations of circular polarizations of excitation and detection. The relative intensities ($I^{++}/I^{+-}/I^{-+}/I^{--}$) of the Mn$^{2+}$ SFRS lines depend strongly on the Mn$^{2+}$ concentration. Here $I^{ij}$ means $\sigma^i$-polarized excitation and $\sigma^j$-polarized detection. These intensities on the Stokes side for $n=1$ at $B=4$~T are given by ($1/0.95/0.99/0.95$) for CdSe/Cd$_{0.991}$Mn$_{0.009}$S (Sample \#1), ($1/0.75/0.84/0.62$) for CdSe/Cd$_{0.99}$Mn$_{0.01}$S (Sample \#2), and ($1/0.33/0.42/0.18$) for CdSe/Cd$_{0.971}$Mn$_{0.029}$S (Sample \#3). One can conclude that the optical selection rules become more distinct with increasing Mn$^{2+}$ concentration and the Mn$^{2+}$ lines become more dominant for $\sigma^{+}$ polarized excitation and detection. 

The observation of multiple spin flip Mn$^{2+}$ lines was reported also for (Cd,Mn)Te-based quantum wells, where up to 15 spin-flip lines were observed.~\cite{Stuhler1995} A mechanism for these flips was suggested in Ref.~\onlinecite{Stuhler1995} and a corresponding model description was developed in Refs.~\onlinecite{Kavokin1997,Merkulov2010ch3}. The key point of this model is the anisotropic spin of the heavy-hole in a two-dimensional nanostructure. In an external magnetic field, the Mn$^{2+}$ spins are polarized along the field direction. When the spin of the photogenerated hole is not parallel to the magnetic field (for simplicity the case when it is perpendicular to the field can be considered), the Mn$^{2+}$ spins are influenced by the external field $\mathbf{B}$ and by the hole exchange field $\mathbf{B}_{\rm exch}$. The total magnetic moment of all Mn$^{2+}$ spins within the hole localization volume, $I_{\rm Mn}$, precesses about the total field $\mathbf{B} + \mathbf{B}_{\rm exch}$. When the exciton recombines, i.e. scatters, the projection of $I_{\rm Mn}$ on $\mathbf{B}$ differs from the initial value by a multiple of the energy of the $n=1$ Mn$^{2+}$ spin-flip. Note that the electron with an isotropic spin does not support this mechanism. Therefore, we can conclude that in the studied DMS NPLs the holes have sufficient wavefunction overlap with the shell Mn$^{2+}$ spins for providing multiple SFRS. This is in line with our conclusions from the DCP data analysis. 

NPLs have a close analogy to quantum wells and the model approach suggested for quantum wells can be directly applied also here. The only specifics, which need to be accounted for, are the varying orientations of the NPLs in an ensemble measurement. As we have noticed above, the condition for observation of multiple Mn$^{2+}$-flips is the noncollinearity of $\mathbf{B}$ and $\mathbf{B}_{\rm exch}$. This means that in quantum wells the effect should be absent in the Faraday geometry, where the magnetic field is parallel to the structure growth axis. In DMS NPLs we observe the same amount of higher order spin scattering resonances in Faraday and Voigt geometry. We explain this result by the random orientation of the NPLs in the studied ensembles, leading to the situation that in any configuration a fraction of NPLs fulfills the condition for multiple Mn$^{2+}$-flips. It is worthwhile to note that the model developed for the exciton, namely for the hole in the exciton, can be readily applied for the negatively charged exciton, as the hole spin acts similar on Mn$^{2+}$ ions in both cases.  

In conclusion, we have demonstrated the exchange interaction of excitons (trions) with the Mn$^{2+}$ ions in CdSe/Cd$_{1-x}$Mn$_x$S core/shell nanoplatelets by means of polarized photoluminescence, optically detected magnetic resonance and spin-flip Raman scattering. One can conclude that these structures can be studied in detail by these experimental approaches that were established for diluted magnetic semiconductors. In particular, assessment of the dynamics for spin-lattice relaxation gives accurate estimates for the Mn$^{2+}$ ion concentration. Our studies may help to functionalize colloidal DMS nanocrystals as magnetic or magneto-optical markers.

\section{Methods}

\textbf{Magneto-optical measurements.} 
The NPLs were dropcasted on a substrate and mounted in a titanium sample holder on top of a three-axis piezo-positioner and placed in the variable temperature insert ($4.2-70$~K) of a liquid helium bath cryostat equipped with a superconducting solenoid (magnetic fields up to 15~T). The measurements were performed in the Faraday geometry (light excitation and detection parallel to the magnetic field direction).
The photoluminescence was excited with a diode laser (photon energy 3.06~eV, wavelength 405~nm) in continuous-wave or pulsed mode (pulse duration 50~ps, pulse repetition rate 500 kHz) with a weak average excitation density of $0.5$ mW$/$cm$^2$.
The PL was dispersed with a 0.5-m spectrometer and detected either by a liquid-nitrogen-cooled charge-coupled-device (CCD) camera or a Si avalanche photodiode connected to a conventional time-correlated single-photon counting system. The instrumental response time was about 200~ps. The PL circular polarization degree was analyzed by a combination of a quarter-wave plate and a linear polarizer. 

\textbf{Spin-flip Raman scattering.} 
The samples were mounted strain free inside the variable temperature insert of a magnet cryostat, which provided magnetic fields up to $10~$T. The temperature was set to $1.6$~K. The backscattering experiments were performed in Faraday geometry ($\theta = 0^{\circ}$) or in tilted geometries up to $\theta=90^{\circ}$, corresponding to the Voigt geometry, where the magnetic field and the normal to the sample substrate enclose the angle $\theta$. The NPLs were excited by a single frequency dye laser (Matisse DS), whose actual wavelength was measured and monitored by a fiber-coupled wavelength-meter device. The laser power was stabilized by a liquid-crystal variable attenuator. Unless specified otherwise, the power was set to about $0.2~$W/cm$^2$ at the sample surface. In order to ensure a stable detection position on the sample surface, each sample was covered by a mask having a hole of 1~mm diameter and the central part with $100\times 100~\mu$m$^2$ size of the illuminated sample area was selected by a cross slit. The NPL emission was spectrally dispersed by a double monochromator (U1000) equipped with a Peltier-cooled GaAs photomultiplier. The SFRS spectra were measured in close vicinity of the laser line with photon energy $E_{exc}$. The spin-flip signals are shifted from the laser energy by the Zeeman splitting of the involved spin state, either to lower energy (Stokes shift, $E_{exc}-|g|\mu_{\rm B} B$) or to higher energies (anti-Stokes shift, $E_{exc}+|g|\mu_{\rm B} B$). 

\textbf{Optically detected magnetic resonance.}
The ODMR technique used in this study was described in detail in Ref.~\onlinecite{Ivanov2007}.  The ODMR spectrometer consisted of a 60~GHz all-solid-state microwave oscillator (photon energy of 0.248~meV) with a tuning range from 59.05 to 60.55~GHz and an output power of up to 100~mW. The output power of the oscillator could be varied by up to 40 dB attenuation level. The oscillator could operate either in continuous-wave mode or in a periodically pulsed mode with an on-off transition time of about 3~ns at more than 60~dB dumping level. The sample was mounted in a cylindrical H$_{011}$ microwave cavity with a low Q factor of about 600. The cavity had two orthogonal pairs of apertures with a conic cross-section for sample illumination and collecting the sample emission. The cavity was placed in the variable temperature insert of a magnet cryostat, the measurements were performed at $T=1.8$~K. The sample in the cavity was excited by a 405~nm ($E_{exc}=2.14$~eV) semiconductor laser with 0.5~mW power, focused into a spot with a diameter of 400~$\mu$m. The photoluminescence was collected in backscattering geometry and detected with a 0.5-meter grating monochromator and a CCD camera. Magnetic fields up to 7~T were applied in the Faraday geometry. For time-resolved ODMR measurements a photon counter based on an avalanche photodiode was used, the temporal resolution was 30 ns.

\textbf{ASSOCIATED CONTENT}

\textbf{Supporting Information.}
Additional information on exciton bright-dark splitting; calculation of the band structure in core/shell NPLs; modeling of the exciton exchange interaction with Mn$^{2+}$ ions, Zeeman splitting and polarization degree for the bright and dark excitons and negative trions; synthesis details.\\

\textbf{AUTHOR INFORMATION}\\
Corresponding Authors:\\
Email: elena.shornikova@tu-dortmund.de\\ 
Email: dmitri.yakovlev@tu-dortmund.de\\

\textbf{ORCID}\\
Elena V. Shornikova: 0000-0002-6866-9013 \\
Danil O. Tolmachev: 0000-0002-7098-8515\\
Vitalii Yu.~Ivanov: 0000-0002-4651-8476\\
Ina~V.~Kalitukha: 0000-0003-2153-6667\\
Victor~F.~Sapega: 0000-0003-3944-7443\\
Dennis~Kudlacik: 0000-0001-5473-8383\\
Dmitri R. Yakovlev: 0000-0001-7349-2745 \\
Yuri~G.~Kusrayev: 0000-0002-3988-6406\\
Aleksandr A. Golovatenko: 0000-0003-2248-3157\\
Shendre Sushant: 0000-0001-8586-7145\\
Savas~Delikanli: 0000-0002-0613-8014\\
Hilmi~Volkan~Demir: 0000-0003-1793-112X\\
Manfred Bayer:  0000-0002-0893-5949\\

\textbf{Notes}

The authors declare no competing financial interests.

\section*{Acknowledgements}
The authors are thankful to  A. V. Rodina for fruitful discussions. This work was supported by the Deutsche Forschungsgemeinschaft through the International Collaborative Research Center TRR 160 (Projects B1 and B2) and by the Russian Foundation for Basic Research (Grant No. 19-52-12064 NNIO-a). D.R.Y. acknowledges the partial support of the Russian Science Foundation (Project No. 20-42-01008). S.S., S.D. and H.V.D. acknowledge partial support from the Singapore National Research Foundation under NRF--NRFI2016--08. A.A.G. acknowledges support of the Grants Council of the President of the Russian Federation. V.Yu.I. acknowledges support of the Polish National Science Center
(Grant No. 2018/30/M/ST3/00276). H.V.D. gratefully acknowledges support from TUBA.

\clearpage

\setcounter{equation}{0}
\setcounter{figure}{0}
\setcounter{table}{0}
\setcounter{page}{1}
\renewcommand{\theequation}{S\arabic{equation}}
\renewcommand{\thefigure}{S\arabic{figure}}
\renewcommand{\thetable}{S\arabic{table}}

\onecolumngrid


\begin{center}
\textbf{\large Supplementary Information:}
	
\vspace{3mm}	
\textbf{\large Magneto-optics of excitons interacting with magnetic ions in CdSe/CdMnS colloidal nanoplatelets}
	
\vspace{3mm}
	
{E.~V.~Shornikova$^1$, D.~R.~Yakovlev$^{1,2}$, D.~O.~Tolmachev$^{1,2}$, V.~Yu.~Ivanov$^3$, I.~V.~Kalitukha$^2$, V.~F.~Sapega$^2$, D.~Kudlacik$^1$,  Yu.~G.~Kusrayev$^2$, A.~A. Golovatenko$^{2}$, S. Shendre$^{4}$, S.~Delikanli$^{4,5}$, H. V.~Demir$^{4,5}$, and M.~Bayer$^{1,2}$}
\end{center}
\vspace{3mm}

{\small \noindent$^1$Experimentelle Physik 2, Technische Universit{\"a}t Dortmund, 44227 Dortmund, Germany
	
\noindent$^2$Ioffe Institute, Russian Academy of Sciences, 194021 St. Petersburg, Russia
	
\noindent$^3$Institute of Physics, Polish Academy of Sciences, PL-02-668 Warsaw, Poland 
	
\noindent$^4$LUMINOUS! Center of Excellence for Semiconductor Lighting and Displays, School of Electrical and Electronic Engineering, School of Physical and Materials Sciences, Nanyang Technological University, 639798 Singapore}

\noindent$^5$Department of Electrical and Electronics Engineering, Department of Physics, UNAM -- Institute of Materials Science and Nanotechnology, Bilkent University, 06800 Ankara, Turkey
		
\vspace{6mm}
\textbf{\large S1. Bright and dark exciton emission}

\begin{figure*}[h!]
	\begin{center}
		\includegraphics{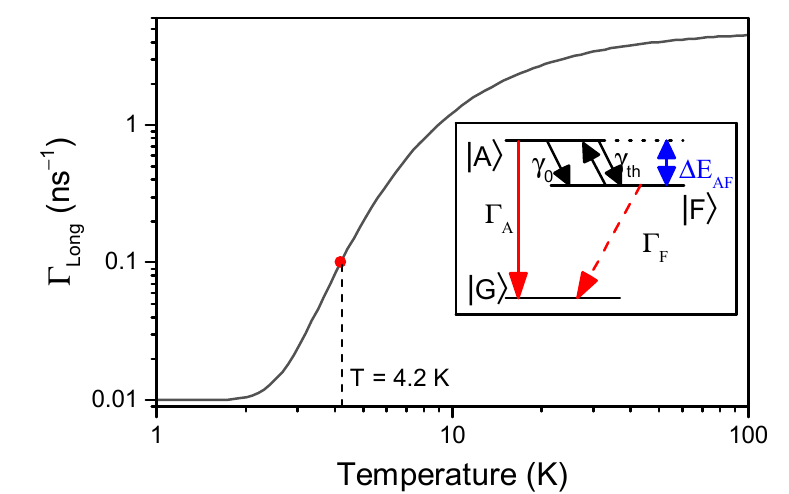}
		\caption{\label{figS1} Long component decay rate $\Gamma_{\rm L} = \tau_{\rm L}^{-1}$ calculated with equation~\eqref{eq:tauLong} as a function of the temperature. The parameters $\Gamma_{\rm A}=10$~ns$^{-1}$ and $\Gamma_{\rm F}=0.012$~ns$^{-1}$ are chosen as in Ref.~\onlinecite{Shornikova2018ns}, despite studying a different sample. $\Delta E_{\rm AF}=1.7$~meV is taken from Table~\ref{tab1}. Inset: Three-level model: $\ket{A}$ and $\ket{F}$ are bright and dark exciton states, and $\ket{G}$ is unexcited crystal state. $\gamma_0$ is zero temperature spin relaxation rate, $\gamma_{th}=\gamma_0N_{\rm B}$ is the thermal-activation rate for the reversed process, where $N_{\rm B} = 1/ \left[ \exp {(\Delta E_{\rm AF} / kT)} -1 \right]$ is the Bose-Einstein phonon occupation. More details in Ref.~\onlinecite{Shornikova2018ns}.}   
	\end{center}
\end{figure*}

In CdSe NPLs the lowest exciton state, $\ket{F}$, has angular momentum projections $\ket{\pm 2}$ from which the emission is forbidden in electric dipole approximation. The optical transitions from the upper-lying $\ket{A}$ state with angular momentum projections $\ket{\pm 1}$ are allowed (inset Fig.~\ref{figS1}). The energy splitting between these two states, $\Delta E_{\rm AF}$, is the bright-dark splitting. The radiative decay rates from these states are $\Gamma_{\rm F}$ and $\Gamma_{\rm A}$, accordingly, and the resulting superposition of decays is bi-exponential with the short decay time below the resolution of our experiment (200 ps), and the long decay time given by
\begin{equation}
\tau_{\rm L}(T)^{-1}=\Gamma_{\rm L}(T)=\frac{\Gamma_{\rm A} +\Gamma_{\rm F}}{2} - \frac{\Gamma_{\rm A}   -\Gamma_{\rm F}}{2}\tanh\left(\frac{\Delta E_{\rm AF}}{2kT}\right) .
\label{eq:tauLong}
\end{equation}
Here we assume that the zero temperature spin relaxation rate $\gamma_0 \gg \Gamma_{\rm A}$. For more details see Ref.~\onlinecite{Shornikova2018ns}.

The $\Delta E_{\rm AF}$ splitting can be measured as the Stokes shift of the emission in fluorescence line narrowing spectra. It equals to 1.7 meV in CdSe/CdS NPLs (Fig.~\ref{fig_SFRS_1}a and Table~\ref{tab1}). Assuming a Boltzmann distribution, this gives the bright exciton population of 1\% at $T=4.2$~K in thermal equilibrium, i.e. bright excitons contribute to the emission. To see, how this affects the decay, we evaluate the average decay time from such system. Since the PL decay is multiexponential, it is not possible to obtain accurate values of $\Gamma_{\rm F}$ and $\Gamma_{\rm A}$. To make an estimate, we use $\Gamma_{\rm A}=10$~ns$^{-1}$ and $\Gamma_{\rm F}=0.01$~ns$^{-1}$, which are reasonable parameters measured for bare core NPLs.~\cite{Shornikova2018ns} The result of the simulation is shown in Fig.~\ref{figS1}. The long component is accelerated by about a factor of 10 when the temperature increases from 1.6~K up to 4.2~K. This might be one of the reasons why the decay is not sensitive to the magnetic field (Fig.~\ref{fig1}d,e). At low temperatures the dark exciton emission is accelerated, because the magnetic field component perpendicular to the quantization axis mixes bright and dark states. In bare CdSe NPLs, the long decay component is 20 times faster at $B=15$~T compared to 0~T.~\cite{Shornikova2020nn} In our case, however, at $T=4.2$~K the decay is already 10 times accelerated due to the thermal population of the bright exciton, which masks the effects caused by the magnetic field. Another reason why the decay is not affected by the magnetic field is the presence of charged excitons (trions).

Note that the ODMR and SFRS experiments were carried out at $T=1.8$ and $1.6$~K, respectively, so that the bright exciton population was negligible.

\vspace{6mm}

\textbf{\large S2. Band structure of CdSe/Cd$_{1-x}$Mn$_x$S NPLs. Electron and hole leakage into shells}

The bandgap difference between the CdSe core and CdS shell equals to 0.75 eV ($E_g^{\rm CdSe}\approx1.75$ eV, $E_g^{\rm CdS}\approx2.5$ eV)\cite{Adachi2004}. Note that the $E_g^{\rm CdSe}=1.84$ eV used in Refs.~\onlinecite{Strassberg2019,Muckel2018} corresponds to wurtzite CdSe, while CdSe nanoplatelets are known to have zincblende crystal structure. 

The conduction and valence band offsets between CdSe and CdS are not precisely known. However, the valence band offsets $\Delta E_v$ reported in the literature are large, at least 0.45~eV, so that the hole is believed to be well confined in the CdSe core. The reported conduction band offsets $\Delta E_c$ range from $0.3$~eV to 0~eV, and this value depends on the crystal structure, NC size, lattice strain and temperature. Due to weak, if present at, confinement, the electron wave function leaks into the CdS shell (for references, see~\cite{Javaux2013}).

\begin{figure*}[h]
	\includegraphics{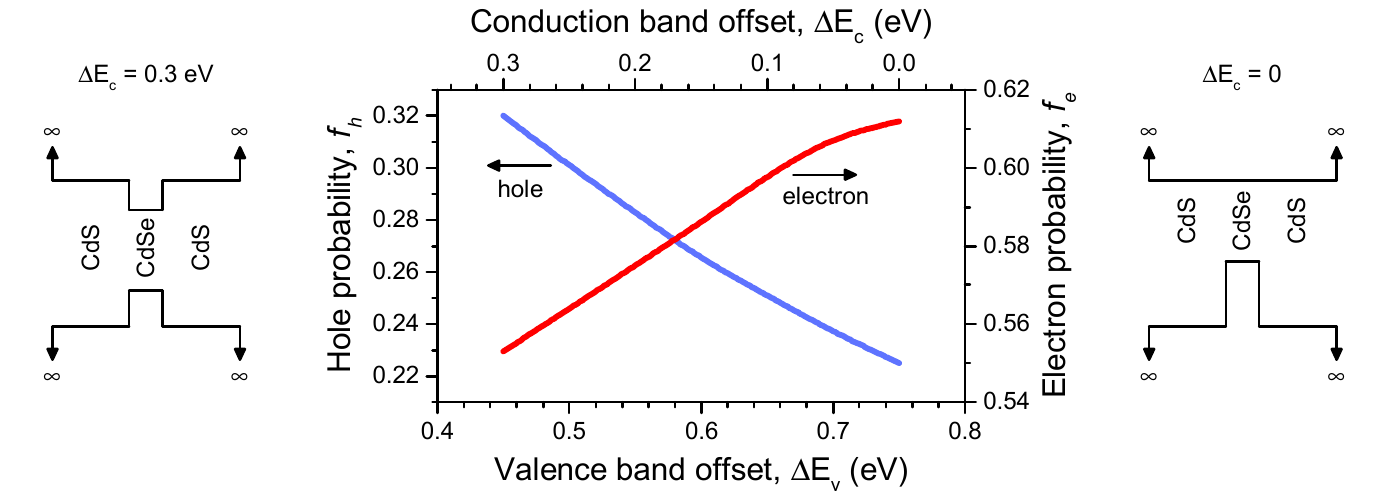}
	\caption{\label{figS2} Electron and hole probabilities for being located in the shell, $f_{\rm e}$ and $f_{\rm h}$, as a function of the conduction and valence band offsets between CdSe and CdS. Schematic band diagrams for $\Delta E_c=0.3$~eV and $\Delta E_c=0$ are shown on the left and on the right, respectively. $f_{\rm e}$ and $f_{\rm h}$ are calculated assuming infinitely high outer barriers.
	}
\end{figure*}

Figure~\ref{figS2} shows the probabilities to find an electron or a hole in the barriers, $f_{\rm e}$ and $f_{\rm h}$, respectively, calculated for NPLs with two monolayer thick (0.6 nm) CdSe cores and four monolayer thick (1.2 nm) CdS shells. The offsets range between two limiting cases shown schematically in Figure~\ref{figS2}: (i) $\Delta E_c=0.3$~eV and $\Delta E_v=0.45$~eV; and (ii) $\Delta E_c=0$~eV and $\Delta E_v=0.75$~eV. For this, we solved the Schr\"{o}dinger equation for the CdSe/CdS system within the effective mass approximation, as it was done in Refs.~\onlinecite{Strassberg2019,Muckel2018}. A conventional set of electron and hole effective masses ($m_{\rm e}$ and $m_{\rm h}$) is used: $m_{\rm e}=0.18m_0$ and $m_{\rm h}=0.89m_0$ in CdSe, $m_{\rm e}=0.35m_0$ and $m_{\rm h}=0.95m_0$ in CdS \cite{Ithurria2011,Strassberg2019,Muckel2018}. Here $m_0$ is the free electron mass. The calculated $f_{\rm h}$ ranges between 32\% and 22\%, while the $f_{\rm e}$ is in the range from 55\% to 61\%. Thus, within the wide range of band offsets, a significant penetration of both electron and hole into the CdS barriers is expected. Note that according to our calculations, the probability to find a hole in the CdS shell is much larger than reported in Refs.~\onlinecite{Strassberg2019,Muckel2018}. Note also that we do not account for the Coulombic attraction between electron and hole, which would provide an attractive potential minimum for the electron centered in the CdSe core even for case of $\Delta E_c=0$~eV.~\cite{Javaux2013}

\vspace{6mm}
\textbf{\large S3. Exchange interaction of excitons and trions with Mn$^{2+}$ ions}

The model descriptions of the DCP for neutral and negatively charged excitons differ, as their Zeeman splittings are determined by the exciton and hole $g$-factors, respectively. In our calculations we use the conventional approach for diluted magnetic semiconductors. This approach is based on consideration of the exchange interaction of an exciton with the Mn$^{2+}$ ions within the mean field approximation.

The bright exciton Zeeman splitting, $\Delta E_{\rm Z,XA}(B)$, in nonmagnetic samples is
\begin{equation} \label{e3-1}
\Delta E_{\rm Z,XA}(B)= g_{\rm XA}\mu_{\rm B}B=(-g_{\rm e}-3g_{\rm h})\mu_{\rm B}B\, ,
\end{equation} 
where $g_{\rm XA}=-g_{\rm e}-3g_{\rm h}$ is the bright exciton $g$-factor, $g_{\rm e}$ and $g_{\rm h}$ are the electron and hole $g$-factors, respectively, and $\mu_{\rm B}$ is the Bohr magneton. In DMS samples an additional term, $E_{\rm exch,XA}(B)$, describing the exciton exchange interaction with the Mn$^{2+}$ spins needs to be added:
\begin{equation} \label{e3-2}
\Delta E_{\rm Z,XA}(B)= g_{\rm XA}\mu_{\rm B}B+E_{\rm exch,XA}(B)=(-g_{\rm e}-3g_{\rm h})\mu_{\rm B}B-\langle S_{\rm Mn}\rangle x\left(N_0\alpha f_{\rm e} - N_0\beta f_{\rm h}\right)\, .
\end{equation}
Here $x$ is the concentration of Mn$^{2+}$ ions, $N_0$ is the number of cations per unit volume, $\alpha$ and $\beta$ are the $s-d$ and $p-d$ exchange constants in Cd$_{1-x}$Mn$_x$S for electron and hole, respectively \cite{Kossut2010}. $\langle S_{\rm Mn}\rangle$ is the mean spin of the Mn$^{2+}$ ions, which depends on the external magnetic field, the temperature and the Mn concentration: $\langle S_{\rm Mn}\rangle=f(x,T,B)$. The latter dependence is provided by the fact that neighboring Mn$^{2+}$ ions interact antiferromagnetically with each other, which reduces $\langle S_{\rm Mn}\rangle$ with increasing $x$.  To account for this, we use the function for the mean spin from Refs.~\onlinecite{Kossut2010,Keller2001}:
\begin{equation} \label{e3-8}
\langle S_{\rm Mn}\rangle=S_0(x)B_{5/2}[g_{\rm Mn}\mu_{\rm B}B/k_{\rm B}(T+T_0(x))]\, ,
\end{equation}
where $B_{5/2}$ is the Brillouin function for spin 5/2, $g_{\rm Mn}=2.01$,  $S_0(x)=-0.804+0.364/(x+0.109)$ is the effective spin, and $T_0(x)=47.2x-281x^2+714x^3$ is the effective temperature. The parameters $S_0(x)$ and $T_0(x)$ describe phenomenologically the interaction between the Mn$^{2+}$ ions.

Note that $E_{\rm exch,XA}(B)$ is controlled by the exchange interaction of both electron and hole composing the exciton with the Mn$^{2+}$ ions. Therefore, it depends on the overlaps, $f_{\rm e}$ and $f_{\rm h}$, of the electron and hole wave functions with (Cd,Mn)S shells, calculated in Section~S2.

The dark exciton Zeeman splitting, $\Delta E_{\rm Z,XF}(B)$, in nonmagnetic samples is
\begin{equation} \label{e3-3}
\Delta E_{\rm Z,XF}(B)= g_{\rm XF}\mu_{\rm B}B=(g_{\rm e}-3g_{\rm h})\mu_{\rm B}B\, ,
\end{equation} 
where $g_{\rm XF}=g_{\rm e}-3g_{\rm h}$ is the dark exciton $g$-factor. In DMS samples, the Zeeman splitting reads as 
\begin{equation} \label{e3-4}
\Delta E_{\rm Z,XF}(B)= g_{\rm XF}\mu_{\rm B}B+E_{\rm exch,XF}(B)=(g_{\rm e}-3g_{\rm h})\mu_{\rm B}B-\langle S_{\rm Mn}\rangle x\left(-N_0\alpha f_{\rm e} - N_0\beta f_{\rm h}\right)\, .
\end{equation}

The Zeeman splitting of the negatively charged exciton, which is composed of two electrons and one hole, is contributed only by the hole:
\begin{equation} \label{e3-5}
\Delta E_{\rm Z,h}(B)= -3g_{\rm h}\mu_{\rm B}B\, .
\end{equation} 
In DMS samples, due to zero total angular momentum of the electrons, the exchange interaction with the Mn-ions does not affect the Zeeman splitting, and $E_{\rm exch,h}(B)$ is determined by the exchange interaction of the hole with the Mn$^{2+}$ ions.
\begin{equation} \label{e3-6}
\Delta E_{\rm Z,h}(B)= -3g_{\rm h}\mu_{\rm B}B+E_{\rm exch,h}(B)=-3g_{\rm h}\mu_{\rm B}B+\langle S_{\rm Mn}\rangle x\left( N_0\beta f_{\rm h}\right)\, .
\end{equation}

\begin{figure*}[h]
	\includegraphics{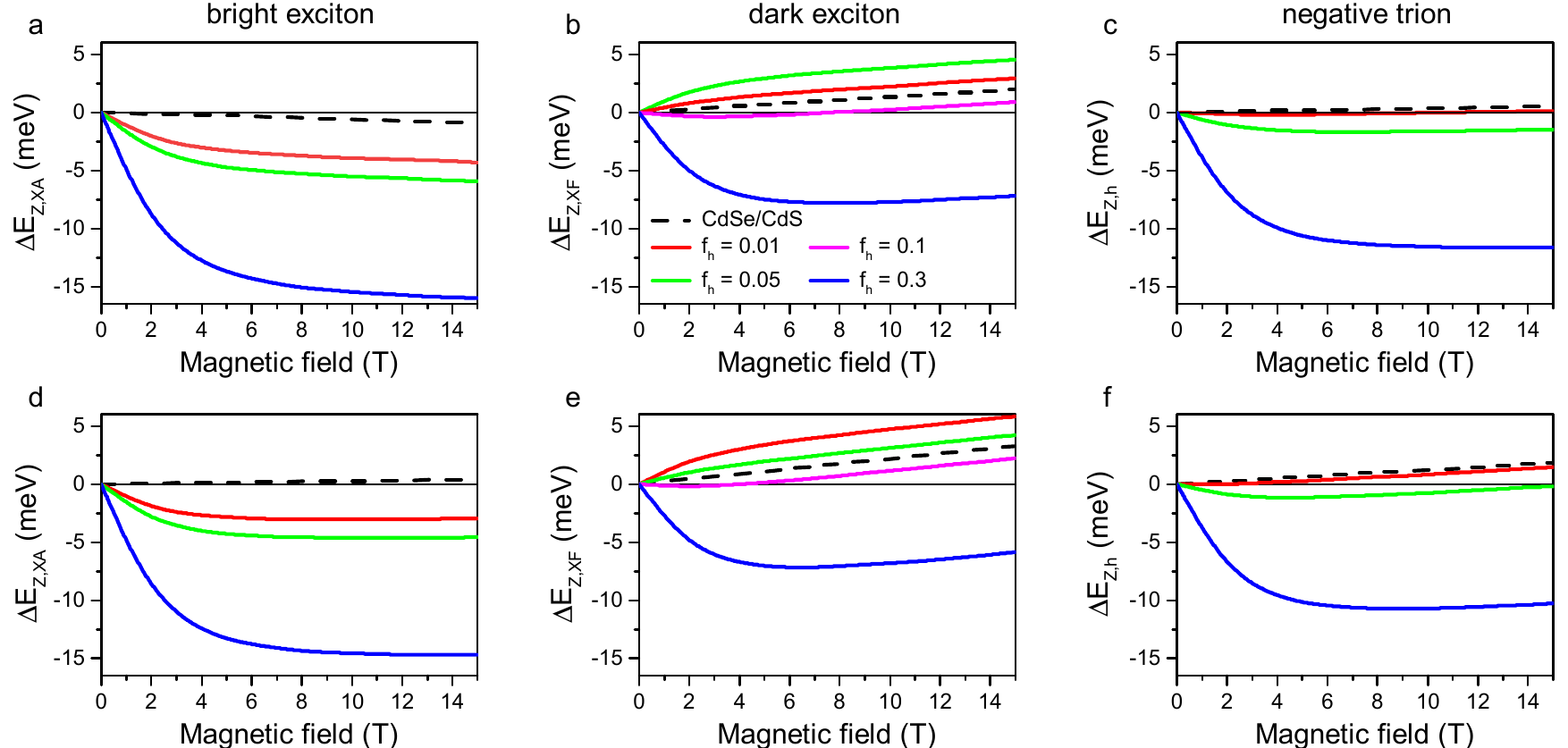}
	\caption{Calculated splittings of the spin sublevels of the bright exciton $\Delta E_{\rm Z,XA}$ (a,d), dark exciton $\Delta E_{\rm Z,XF}$ (b,e) and negative trion $\Delta E_{\rm Z,h}$ (c,f) vs magnetic field. In panels (a--c) the hole g-factor $g_{\rm h}=-0.2$, while in panels (d--f) $g_{\rm h}=-0.7$. The black dashed lines correspond to the splitting in nonmagnetic CdSe/CdS NPLs. The red, green, pink and blue lines correspond to CdSe/Cd$_{0.991}$Mn$_{0.009}$S NPLs with $f_{\rm e}=0.6$ and different $f_{\rm h}$ shown in panel (b).}
	\label{figS3}
\end{figure*}

Figure~\ref{figS3} shows the Zeeman splittings of the bright excitons (a,d), dark excitons (b,e) and negatively charged excitons (c,d) as a function of the magnetic field. The following parameters were used for the calculation: $g_{\rm e}=1.7$ (determined from spin-flip Raman scattering), $N_0\alpha=0.22$~eV and $N_0\beta=-1.8$~eV \cite{Kossut2010}, $S_0(x)=2.25$ and $T_0(x)=0.45$. The calculations were done for two hole $g$-factors: the theoretically calculated $g_{\rm h}=-0.2$ (a--c), and the experimentally measured $g_{\rm h}=-0.7$ (d--e), both values taken from Ref.~\onlinecite{Shornikova2018nl}. Note that in Ref.~\onlinecite{Shornikova2018nl} the hole $g$-factor increased with the magnetic field strength from $g_{\rm h}=-0.4$ in $B=0$ to $g_{\rm h}=-0.7$ in $B=15$~T due to band mixing effects. Here we use the high magnetic field value.  

The exciton and trion DCP is given by
\begin{equation} \label{e3-7}
P_{\rm c}(B)= -P^{sat}_{\rm c} \frac{\tau}{\tau+\tau_{s}} \tanh  \frac{\Delta E_{\rm Z}(B)}{2k_{\rm B}T} \, .
\end{equation} 
Here $\tau$ and $\tau_{s}$ are the lifetime and spin relaxation time, respectively, $P^{sat}_{\rm c}$ is the saturation degree, which depends on specifics of the spin structure and NPL orientation in the ensemble.

\begin{figure*}[h!]
	\includegraphics{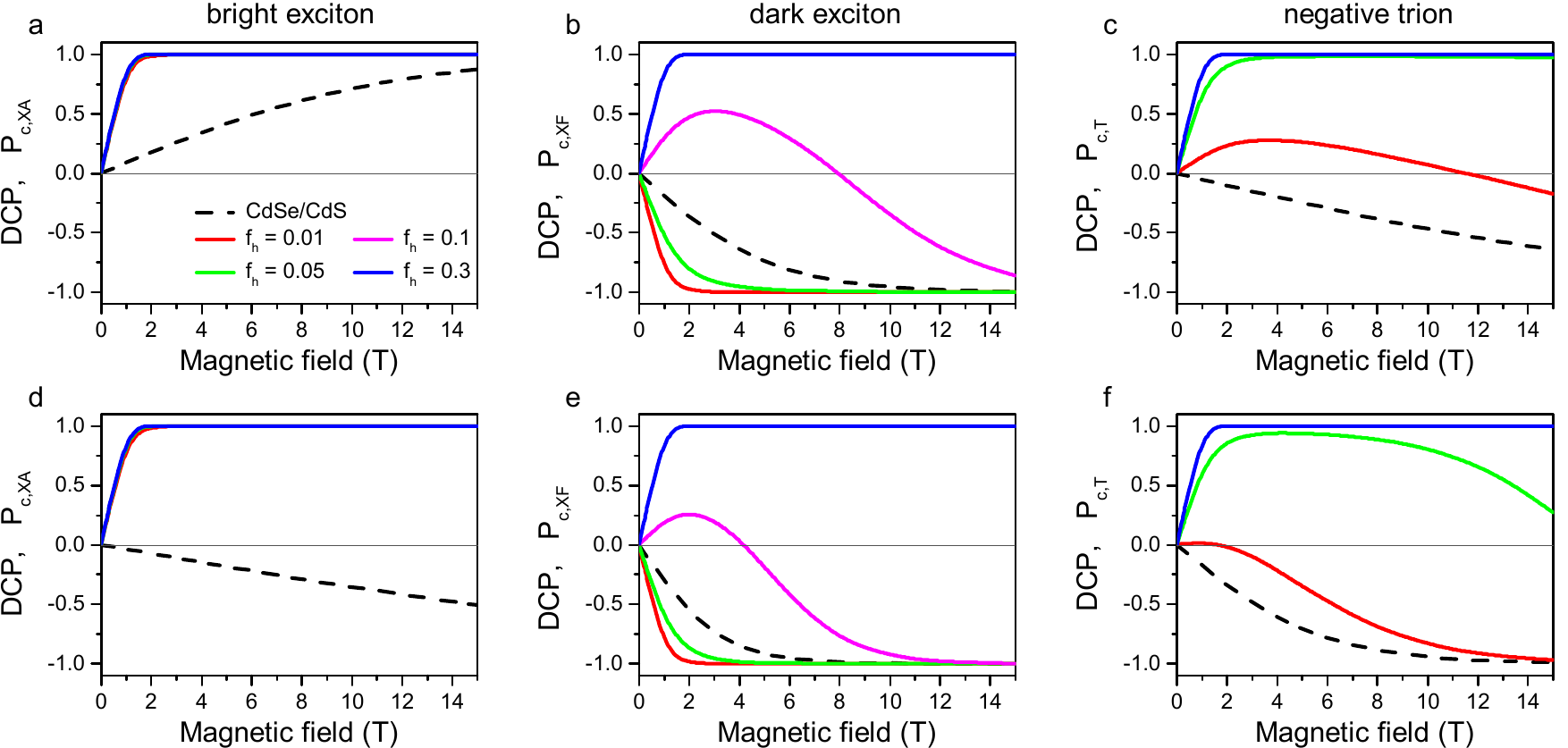}
	\caption{Calculated DCP for the bright exciton $P_{\rm c,XA}$ (a,d), dark exciton $P_{\rm c,XF}$ (b,e) and negative trion $P_{\rm c,T}$ (c,f) for $T=4$~K as a function of the magnetic field. In panels (a--c) hole g-factor $g_{\rm h}=-0.2$, while in panels (d--f) $g_{\rm h}=-0.7$. The black dashed lines correspond to the polarization in nonmagnetic CdSe/CdS NPLs. The red, green, pink and blue lines correspond to CdSe/Cd$_{0.991}$Mn$_{0.009}$S NPLs with $f_{\rm e}=0.6$ and different $f_{\rm h}$ shown in panel (a).}
	\label{figS4}
\end{figure*}

Figure \ref{figS4} shows DCP vs magnetic field plots based on the Zeeman splittings from Figure~\ref{figS3}. They were calculated for $P^{sat}_{\rm c}=1$, $\tau \gg \tau_{s}$, and $T=4$~K, assuming a Boltzmann distribution between the spin sublevels and neglecting the interaction between bright and dark excitons, i.e. they can be treated as DCP when only one type of particles -- bright or dark excitons, or trions -- is present in the system. However, we note that due to the small $\Delta E_{AF}\approx1.7$~meV in the samples under investigation the exchange interaction of carriers with the Mn$^{2+}$ ions will result in a crossing of the bright and dark exciton energy levels already in low magnetic fields, which requires further study.  

In nonmagnetic CdSe/CdS NPLs (the black dashed lines in Figures~\ref{figS3} and \ref{figS4}), a positive Zeeman splitting, i.e. a negative DCP, is expected for the dark excitons and negative trions. The $P_{\rm c}$ sign of the bright excitons depends on the hole $g$-factor. It is negative when $g_{\rm h}=-0.7$, and positive when $g_{\rm h}=-0.2$. The experimentally measured $P_{\rm c}<0$, but we cannot make a definite conclusion about the $g_{\rm h}$ value, because in experiment all three exciton complex species are present in the spectra.

In DMS NPLs (the color lines in Figures~\ref{figS3} and \ref{figS4}), $P_{\rm c}>0$ is expected for all species, provided that the hole leakage into the DMS shells is large enough. For $g_{\rm h}=-0.2$, this requires at least $f_{\rm h}>10\%$ to change the DCP sign of the dark excitons, $f_{\rm h}>5\%$ for the negative trions, and an even larger $f_{\rm h}$ in case $g_{\rm h}=-0.7$. As discussed in Section~S2, we estimate $f_{\rm h}$ to be between 22\% and 32\%, i.e. it is even larger than required.

Note that if $f_{\rm h}<5\%$, a DCP sign reversal in DMS NPLs can be provided only by the bright excitons, and only when $g_{\rm h}=-0.7$ (Figure~\ref{figS4}). But this requires that the bright excitons are the main species, which contributes to the emission. This contradicts the average lifetime, which would be very short in this case (below 1 ns).

\vspace{6mm}
\textbf{\large S4. Sample preparation}

\textit{Chemicals:} Cadmium acetate dihydrate ($\text{Cd(OAc)}_2\cdot 2\text{H}_2\text{O}$), trioctylamine (TOA), oleylamine (OLA), N-methylformamide (NMF), ammonium sulfide solution (40-48 wt. \% in water), trioctylphosphine (TOP), oleic acid (OA), hexane, acetonitrile, toluene and manganese(II) acetate were bought from Sigma-aldrich.

\textit{Synthesis of 2 ML CdSe Nanoplatelets:} The synthesis of the NPLs was performed according to a previously reported method.~\cite{Delikanli2019} The mixture of 860 mg $\text{Cd(OAc)}_2\cdot 2\text{H}_2\text{O}$, 1 mL of OA, and 60 mL of TOA was degassed for 1 h at room temperature. Then, it was heated to 115$^\circ$C under argon flow. When the temperature reached 115$^\circ$C, 1 mL of 1M TOP-Se was injected swiftly and the mixture was kept at 115$^\circ$C for 2 h. After that, the solution was cooled down to room temperature and centrifuged after addition of ethanol and hexane. The precipitated NPLs were dispersed in hexane.

\textit{Synthesis of CdSe/CdMnS core/shell NPLs:} Here we used a modified procedure of the c-ALD recipe reported previously.~\cite{Shendre2019} 2 ML CdSe NPLs were dispersed in 1 mL hexane and 5 mL of NMF with 40 $\mu$L of 40--48\% aqueous solution of ammonium sulfide -- as sulfur shell growth precursor -- was added on top of the NPL dispersion and stirred for 2 min. Then, the reaction was stopped by addition of acetonitrile and excess toluene and the mixture was precipitated via centrifugation. The precipitate was redispersed in NMF and precipitated again after addition of acetonitrile and toluene to remove the unreacted precursor. Finally, the NPLs were dispersed in 4 mL of NMF. The cation precursor solution consists of $\text{Mn(OAc)}_2$ and $\text{Cd(OAc)}_2\cdot 2\text{H}_2\text{O}$ in NMF. For the cation deposition step, 1 mL of cation precursor mixture was added to the NPL dispersion and it was stirred for 45 min in a nitrogen filled glovebox. Then, the reaction was stopped by addition of excess toluene and the mixture was precipitated via centrifugation and dispersed in NMF. The same cleaning step was repeated twice more to remove the excess precursors. To increase the number of shells, the steps explained above were repeated until the desired shell thickness was achieved. Lastly, 5 mL of hexane and 100 $\mu$L of OLA were added on top of the precipitated NPLs after achievement of the desired shell thickness and the mixture was stirred overnight. To remove the excess ligands, the dispersion of NPLs was precipitated by addition of ethanol, redispersed and kept in hexane for further usage. The doping levels were obtained using ICP-MS measurements and by taking into account the 2D planar geometry of the NPLs.

\end{document}